\documentclass[11pt,onecolumn]{article}
\usepackage{graphicx}
\usepackage{amssymb,amsfonts,amsmath,array}
\usepackage{authblk}
\usepackage{cite}
\usepackage{xcolor}
\usepackage{comment}
\usepackage{multirow}
\usepackage[sectionbib,square,comma,numbers]{natbib}
\usepackage[outdir=./]{epstopdf}
\usepackage{caption}
\usepackage{subcaption}

\title{\vspace{-2.0cm}
GLN -- a method to reveal unique properties of lasso type topology in proteins}

\author[1,2]{Wanda Niemyska}
\author[3]{Kenneth C. Millett}
\author[2]{Joanna I. Sulkowska\thanks{jsulkowska@cent.uw.edu.pl}}
\affil[1]{Faculty of Mathematics, Informatics and Mechanics, University of Warsaw, Banacha 2, 02-097 Warsaw, Poland}
\affil[2]{Centre of New Technologies, University of Warsaw, Banacha 2c, 02-097 Warsaw, Poland}
\affil[3]{Department of Mathematics, University of California Santa Barbara, CA 93106, USA}

\date{}
\begin{document}
\maketitle  


\begin{abstract}
Geometry and topology are the main factors that determine the functional properties of proteins. In this work, we show how to use the Gauss linking integral (GLN) in the form of a matrix diagram -- for a pair of a loop and a tail -- to study both the geometry and topology of proteins with closed loops e.g. lassos. We show that the GLN method is a significantly faster technique to detect entanglement in lasso proteins in comparison with other methods. Based on the GLN technique, we conduct comprehensive analysis of all proteins deposited in the PDB and compare it to the statistical properties of the polymers. We found that there are significantly more lassos with negative crossings than those with positive ones in proteins, the average value of maxGLN (maximal GLN between loop and pieces of tail) depends logarithmically on the length of a tail similarly as in the polymers. Next, we show the how high and low GLN values correlate with the internal flexibility of proteins, and how the GLN in the form of a matrix diagram can be used to study folding and unfolding routes. Finally, we discuss how the GLN method can be applied to study entanglement between two structures none of which are closed loops. Since this approach is much faster than other linking invariants, the next step will be evaluation of lassos in much longer molecules such as RNA or loops in a single chromosome. 
\end{abstract}


\section*{Introduction}

The protein backbone describes a collection of space curves, a type of spatial structure that mathematicians have been analysing and comparing for a long time. One well-known measure of how two such curves interact with one another is the Gauss linking integral, which is related to Ampere's law of electrostatics and has important applications in modern physics. For two oriented closed curves the Gauss linking integral is always integer, called the linking number, giving an integer invariant describing the number of times one curve winds around the other. The linking number of two not linked curves is $0$, while the Hopf link is the simplest link with linking number equal to +1 or -1, depending upon the relative orientation of the curves \cite{degr}, see Supplementary Information Fig.~1. 

Protein chains are open curves which is often challenging for mathematicians, and induces high computational complexity of algorithms involving randomness and statistics \cite{virnau2006intricate,millett2013identifying}, as in the case of identifying knots \citep{knotdet}, slipknots \citep{king2007identification,sulkowska2012conservation} and links in proteins \citep{dabrowski2017topological}. Against such a backdrop, the fact that Gauss linking integral may be defined generally for open curves and calculated precisely for polygonal chains makes this measure particularly attractive.

The first biological applications of the Gauss linking integral are found in studies of DNA structure \citep{white1969self}.
In 2002, R{\o}gen and Fain applied this measure for comparing and effective classifying protein structures~\citep{rogen2003automatic}. More recently, the Gauss integral has been used for identifying linking in domain-swapped protein dimers~\citep{baiesi2017exploring}. 

\begin{figure}[htb]
\begin{center}
\includegraphics[width=0.8\textwidth]{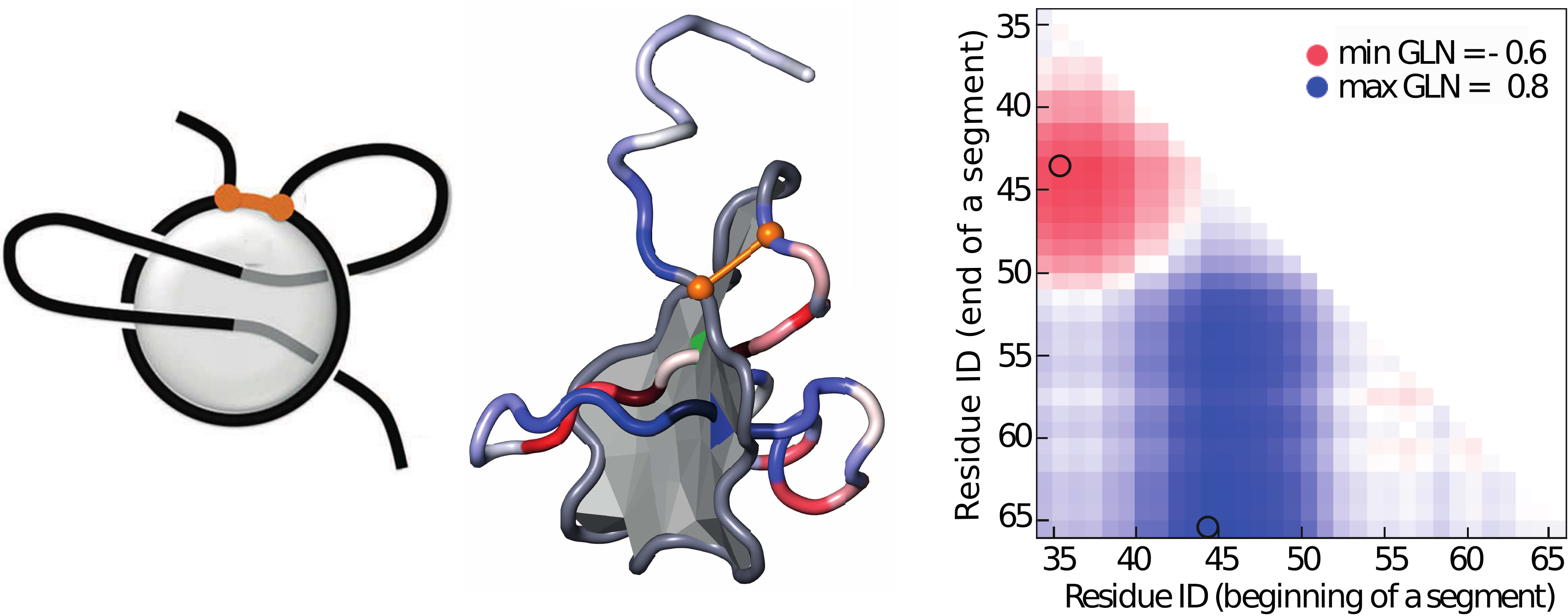}
\caption{Left panel: An example of a lasso configuration of $L_2$ type, with a disulfide bridge (in orange) closing a covalent loop, and a minimal surface (in gray) which spans the loop and is pierced twice by the tail. Middle panel: A cartoon representation of a hydrolase protein (PDB code 5uiw, chain B), with disulfide bridge between amino acids 10 and 34.  It is of $L_2$ type, with minimal surface (in gray) and tails coloured according to the GLN values between their segments and whole loop. Right panel: The topological fingerprint of a lasso based on the GLN matrix for the same protein. Each cell of the matrix corresponds to the GLN value between the disulfide loop and the specific subchain of the tail (here C terminus, the longer one), where the id of the first residue is on the x-axis and the id of the last residue is on the y-axis, thus the left bottom corner corresponds to the whole tail. The C-tail in the middle panel is colored according to the diagonal of the matrix.}
\label{main}
\end{center}
\end{figure}

In this paper we show that the Gauss linking integral, which we denote by GLN, captures unique properties of lasso proteins 
(Fig.~\ref{main}), another type of non-trivial topology identified recently in proteins containing a disulfide or other type of bridge \citep{niemyska2016complex,dabrowski2016lassoprot}. Complex lasso topology is found in at least 18\% of all proteins with disulfide bridges in a non-redundant subset of PDB, and thus represents the largest group of proteins with non-trivial topology.

Lassos occur in structures with disulfide (or other) bridges creating a loop and a pair of termini.  When at least one terminus
of a protein backbone is entangled with the covalent loop (closed by such a bridge) a topologically complex structure is formed.
The topology is identified by a spanning specific surface (i.e. minimal surface) on the covalent loop (Fig.~\ref{main}) and identifying  the crossings of the tails and the surface \citep{niemyska2016complex}. Currently several classes of lasso structures in proteins are known. In addition to the trivial lasso L$_0$, the principal structures are the single lasso L$_1$,  the double lasso L$_2$, and the triple lasso L$_3$, depending upon whether the loop is pierced once, twice and three times, respectively, by the same tail, which goes through the loop and turns back several times. The structure with more than one piercing from the same direction is called a lasso supercoiling LS (when one tail pierces the loop then winds around the protein chain comprising the loop and pierces it again).  Another case identified in proteins is the two-sided lasso LL (when a loop is pierced by both tails). It is important to note that from mathematical point of view all classes of lassos are topologically equivalent to trivial lasso L$_0$ because the free ends are not prevented from unwinding. 
And even if we connected free ends not disturbing windings, except lasso
supercoil LS the rest would be still topologically equivalent to trivial lasso.
But, from biological point of view, they are still very interesting complex structures. For example, a correlation between a type of lasso topology and the specific function of protein has been identified \citep{niemyska2016complex}. 
All proteins that form any type of lasso are collected in the LassoProt database \citep{dabrowski2016lassoprot}.  

Proteins with lassos are found in all domains of life and possess diverse functions \citep{niemyska2016complex,dabrowski2016lassoprot}. Lasso topology can influence thermodynamics properties and biological activity of proteins \citep{haglund2012unique,haglund2014pierced}. 
Cystein bridges provide stability to protein structures and a non-trivial topology can enhance this influence \citep{dabrowski2017topological,niewieczerzal2019supercoiling}. However, it is also known that non-trivial topology hinders the folding pathway \citep{sulkowska2009dodging}, leading to possible misfolding \citep{qin2015protein}. 
How evolution solves this delicate balance is one of the open questions.  There are many others at the interface of biology and mathematics. What is the role of the lasso?  Is there a correlation between the lasso type and the biological function? How do these proteins fold in oxidative conditions? The latter question however does not concern the lasso peptides which are class of ribosomally synthesized posttranslationally modified natural products found in bacteria. However these peptides have a diverse set of pharmacologically relevant activities, including inhibition of bacterial growth, receptor antagonism, and enzyme inhibition \citep{maksimov2012precursor}. Thus, can lasso topology be useful in bioengineering or in pharmacological applications to design proteins with desired fold, stability or other features? 
In polymer chemistry, lassos (known as tadpoles) are used to design materials with desired properties \citep{tezuka2001topological,kricheldorf2010cyclic,tezuka2017topological}.  
Since lassos are defined using open curves they are also inspiring mathematicians to construct topological tools 
capable of classifying them \citep{tian2017knot,nawiasy}. However, up to now, the question of whether a loop and a tail can be entangled in protein while the minimal surface spanned on the loop is not pierced, hasn't been asked.  How might this entanglement influence protein biophysical properties? The Gauss linking integral approach could reveal more information about lasso proteins than the previous geometric method.

The aim of this research is to better understand the entanglement of lasso proteins and its influence on their thermodynamical properties. To do so we first introduce a new technique based on the Gauss linking integral and, then, apply it to assess the topological complexity of proteins with disulfide bridges. We show that GLN provides new information about the entanglement of the loop and tails, related to geometric features of the minimal disc piercings but, in addition, identifies entangled proteins with different complex lasso topology.
We introduce GLN fingerprint to display the local winding of a protein backbone and as another method to quantify entanglement in proteins with non-trivial linking topology. Finally, we use GLN as descriptor to study the free energy landscape of proteins and show influence of non-trivial topology on proteins stability and folding pathway. 




\section*{Results}

Our new approach relies on the definition of the Guass linking integral. 
Let us first consider a protein chain with a disulfide bond connecting two amino acids that, in this way, creates an unknotted covalent loop. The complementary parts of the chain are the tails. When at least one tail pierces a minimal surface spanned on the loop, the entire structure is called a complex lasso (Fig.~\ref{main}). 
In this study, we compute the Gauss linking integral, which we denote by GLN, quantifying the linking  between each tail and the closed loop. The GLN is an algebraic measure of how many times (and in which direction) the tail winds around the loop, with cancellation. For example, a value of GLN close to $1$ means that the tail winds around the loop more or less once, in total. In the most simple cases, the tail passes once through the surface spanned on the loop (in a positive direction, following natural orientation of protein from the N terminus to C terminus).  Such structure resembles the single lasso called L$_1$.  If the direction is reversed, the linking number is close to $-1$. Note that, in complex cases, the tail can pass around the loop twice in a positive direction and once in a negative direction for an algebraic total of about $1$. Moreover, by definition, the linking number of two unlinked curves is 0 although one can not infer with certainty that linking number 0 curves can be separated. This is demonstrated by the ``Whitehead'' link in which the algebraic linking of the two closed loops is zero but they are geometrically entangled and one chain intersects a minimal surface spanned on the other chain at least twice in opposite and therefore cancelling directions. We will present conditions to identify and classify proteins with cystein bridges. 

\subsection*{GLN definition from protein perspective}

The mathematical definition of linking number between two closed curves in $3$ dimensions is given by the Gauss double integral. In the case of proteins, the molecular chains become collections of points, i.e., positions of C$\alpha$ atoms, and the integrals may be replaced by sums of exact quantities determined by pairs of segments connecting the points as determined by the molecular chain \citep{banchoff}. We must relax the expectation of having an integer indicator of linking as we perform the double Gauss integral over open chains. See the section Materials and Methods for the details.
We propose the analysis of four main values for each pair consisting of a loop and a tail: \vspace{0,2cm}\\
$1)~whGLN$: the GLN value of a loop and a whole tail, \\
$2)~minGLN$, and\\
 $3)~maxGLN$\\ respectively, the  minimum and maximum values of GLN between a loop and any fragment of a tail, and\\
$4)~max|GLN|=\max\{maxGLN,-minGLN\}$. \vspace{0,2cm}\\
Additionally, for each triple of a loop and two tails, we consider $max2|GLN|$ value defined to be the maximum of $max|GLN|$ values for both tails. We determine the positive directions of windings according to natural direction of a protein chain; oriented from the $N$-terminus to the $C$-terminus. A high $maxGLN$ or low $minGLN$ indicate that the corresponding part of a tail significantly winds around a loop in a "positive" or "negative" direction, respectively.  Usually the minimal surface spanned on the loop is pierced by this part of the tail. 


We analyzed the entire set of all 5,106 non-redundant proteins in the Protein Data Bank with at least one disulfide bridge (13,320 covalent loops in a total) – from the LassoProt database~\citep{dabrowski2016lassoprot}.  See Materials and Methods section for the details about the dataset. 

\begin{figure}[htb]
\begin{center}
\includegraphics[width=0.5\textwidth]{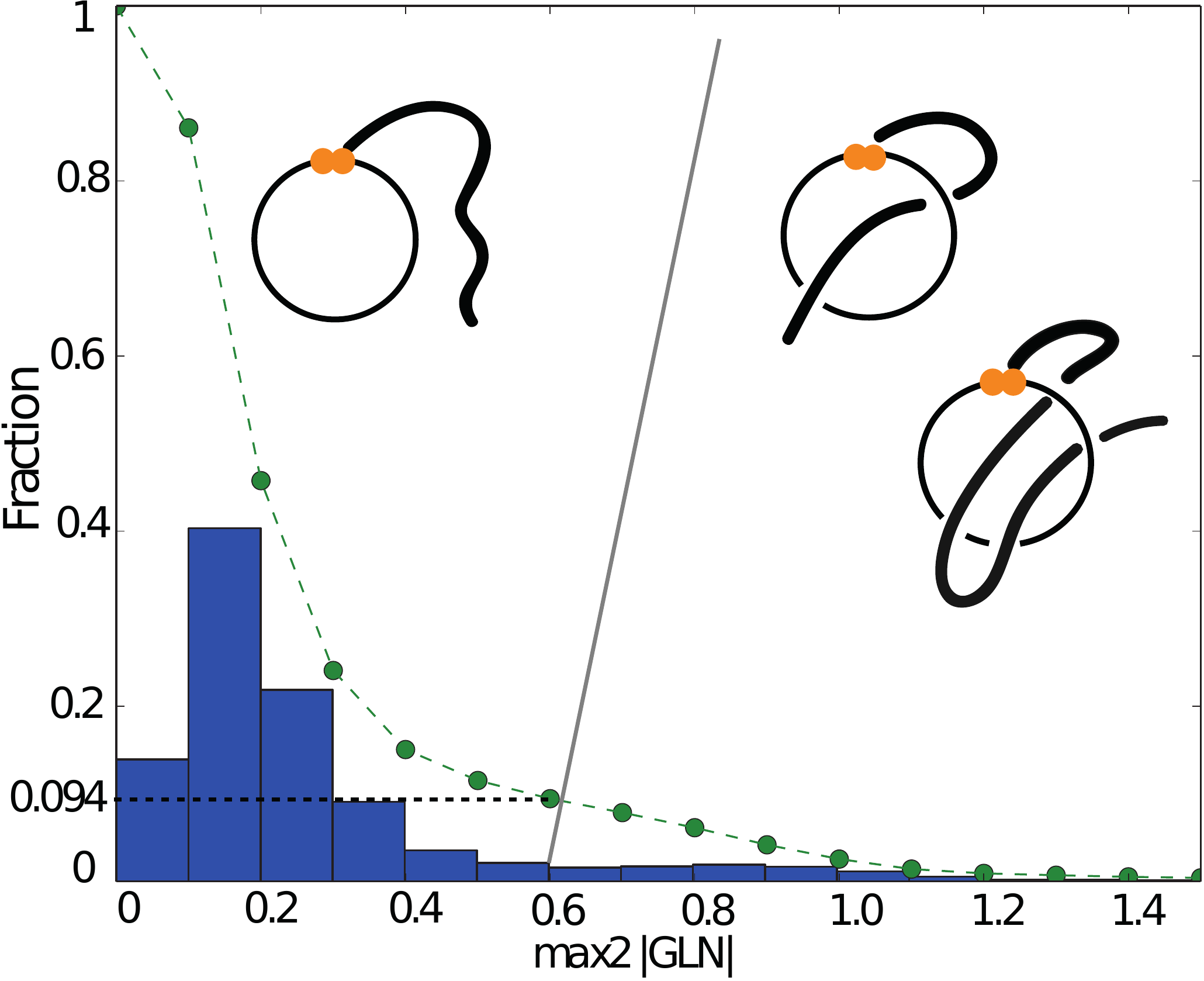}
\caption{The histogram of $max2|GLN|$ values for all closed loops (created by a disulfide bridge) in the set of 5106 non-redudant proteins. The dotted curve shows the fraction of loops having $max2|GLN|$ greater than the value on the x-axis. Almost $10\%$ of the loops have $max2|GLN|$ greater than $0.6$ indicating significant entanglement with a tail. Schematic figures show the most probable corresponding type of the lasso structure. }
\label{fig-much}
\end{center}
\end{figure}

Application of GLN to this dataset reveals the gaussian distribution with long tail as shown in Fig.~\ref{fig-much}. In the majority of cases, the GLN is near 0.2 indicating proteins in which t the minimal surface spanned on the loop is probably not pierced. However, the long tail shows that, in high fraction of chains with cysteine bridges at least one tail significantly winds around the loop. For example, in $21\%$ of chains, we have at least one loop with $max2|GLN|>0.6$ and, in $9.4\%$ of loops, we have $max2|GLN|>0.6$. The value $0.6$ seems to be a good threshold with which to distinguish between complex and trivial topologies, since over $93\%$ of loops with $max2|GLN|>0.6$ have the minimal surface spanned on the loop pierced by a tail at least once and only $4\%$ of loops with $max2|GLN|\leq0.6$ have loop spanning surfaces pierced by either tail.

\subsection*{The GLN fingerprint as a method to classify lasso structures}

To identify the correlation between topology and geometry of proteins, we adopt the idea of topological fingerprint used to exhibit the internal knots in proteins called slipknots \citep{yeates2007knotted,sulkowska2012conservation}.  Here, we present the linking complexity in the form of a matrix diagram -- for a pair of a loop and a tail -- that shows the GLN between the loop and the entire tail and each of its subchains. \\

\begin{figure*}
\centering
  \includegraphics[width=.95\linewidth]{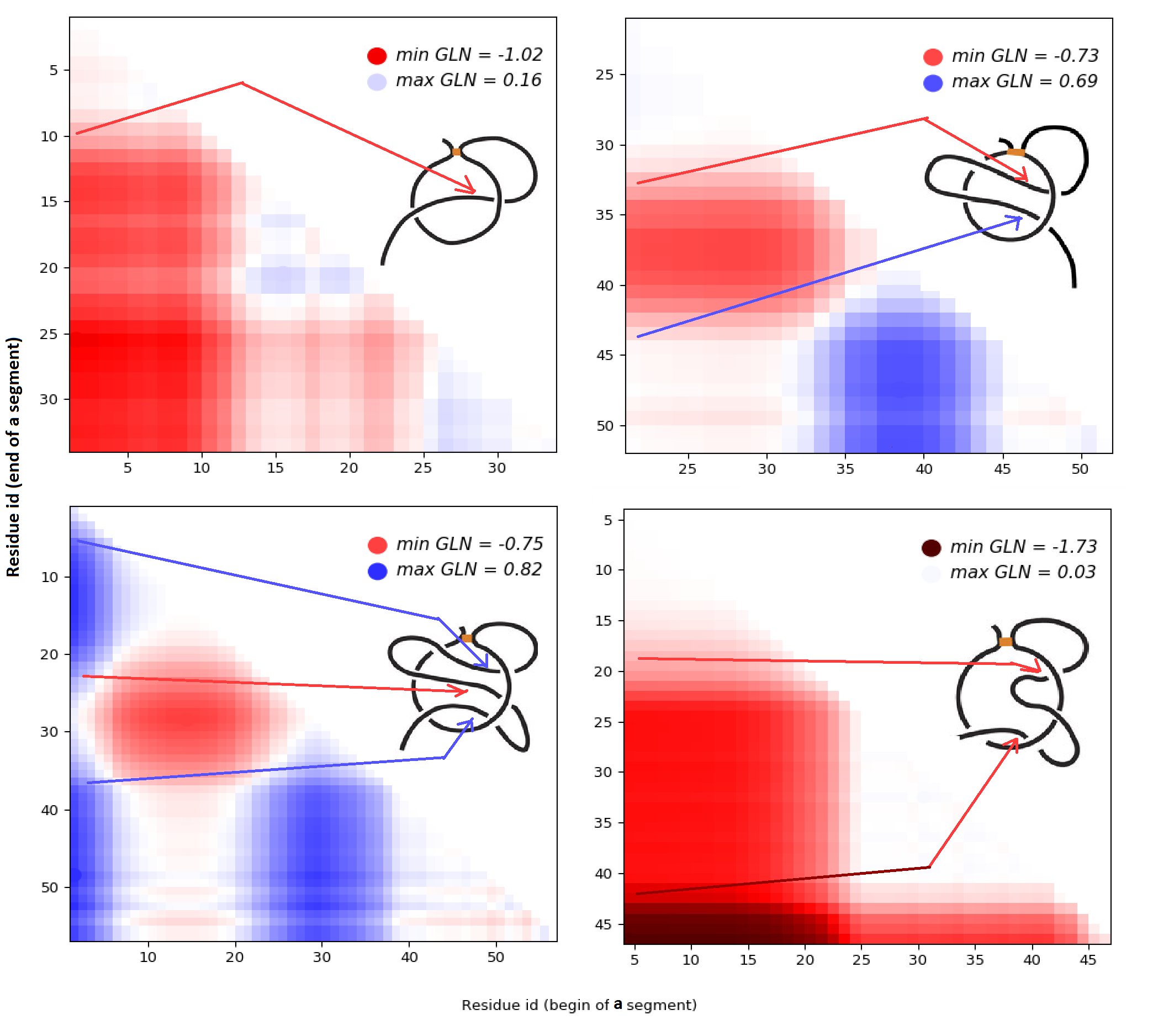}
  \caption{Topological fingerprints -- GLN matrices. Left, the fingerprints gL$_1$ (top) and gL$_3$ (bottom), respectively, for proteins with one and three piercings of the , based on proteins with pdb codes 1i1j and 2ehg. Right, the fingerprints gL$_2$ (top) and gLS (bottom), respectively, for proteins with two piercings of the minimal surface spanned on the loop in the opposite direction and the same direction (supercoiling), based on proteins with pdb codes 2ehg and 1zd0. Arrows begin in the places on the matrices where color is rapidly changing implying that the tail is in the critical phase of winding around the loop and the GLN is quickly increasing or decreasing.  On the other side, on the diagrams they indicate the neighborhoods of possible corresponding piercings. The colors of the arrows indicate directions of windings. }
  \label{fig:matrices}
\end{figure*}

The analysis of our dataset reveals that covalent loops in proteins can be classified into a few distinct motifs, represented by particular patterns within the matrix diagrams. Four characteristic motifs are shown in Fig.~\ref{fig:matrices}. Each point of the matrix corresponds to a specific subchain of the tail, where the id of the first residue is on the x-axis and the id of the last residue is on the y-axis.  As a consequence, the left bottom corner corresponds to the whole tail. The color intensity indicates the value of the GLN between the disulfide loop and the specific subchain of the tail.
A red color indicates negative linking values reflecting the negative direction while blue indicates positive linking values. 
These GLN matrices are used to introduce the following classification of proteins with cystein bridges: \\
$\bullet$ {\bf gL$_0$}, no clear colorfull patches in the matrix indicating that the tail does not wind around the loop. \\
$\bullet$ {\bf gL$_1$}, there is one colorfull patch in the matrix (e.g. in the left bottom corner) indicating that the tail winds around the loop once.  The color indicates the direction. \\
$\bullet$ {\bf gL$_2$}, there are two patches in different colors in the matrix, (e.g. one on the left edge and second one on the bottom edge).  This indicates that the tail winds around the loop in one direction and then in the opposite direction. 
(This spatial arrangement can be observed by following the left edge of the matrix in a descending direction: the beginning of the analyzed segment remains the same - beginning of the tail - while the end of the analyzed segment is moving towards the end of the tail.  When we approach the patch, a color begins to appear meaning the tail begins to wind around the loop. Below the colorfull patch we again see white indicating that the tail winds around the loop but in the opposite direction thereby cancelling the initial winding contribution.  Thus the windings "cancel" themselves and the corner of matrix is again almost white (see Fig.~\ref{main}).). 
\\
$\bullet$ {\bf gL$_3$}, there are four colorfull patches in the matrix, e.g. one in the middle in the different color than three other patches; this indicates that the tail winds around the loop in one direction, then turns and winds around the loop in the opposite direction, and finally turns back one more time. \\
$\bullet$ {\bf gL$_n$}, for any natural $n$, there is specific, dependent on $n$, number of colorfull patches (namely $\left\lfloor \frac{n+1}{2} \right\rfloor\cdot\left\lfloor \frac{n+2}{2} \right\rfloor$) in the matrix; this indicates that tail winds around the loop $n$ times, each next time in the opposite direction. \\
$\bullet$ {\bf gLS}, there is usually one big patch in one color which at some point becomes very intensive - claret or navy in the case of negative and positive windings, respectively; this means that the tail winds around the loop in one direction (making a full circle) and then winds around it one more time in the same direction. \\
$\bullet$ {\bf gLL}, if both matrices for two tails have at least one colorfull patch; this indicates that both tails wind around the loop. \\

Similar GLN matrices indicate the same topological motifs even though the chains may have a different structure. Examples of the same GLN matrices for proteins with very low sequence similarity are shown in Supplementary Information (Fig.~4 and~Fig.~5). The motifs gL$_n$, gLS and gLL usually correspond to the lasso types L$_n$, LS and LL, respectively.  The GLN matrices reveal much more detail about the geometry of the chains with lassos.   By analysing the location, size and color of a collection of  patches one may deduce which parts of the tail wind around the loop and how fast and tightly they wind. For the most part intense patches correspond to the tail piercing 
the minimal surface spanned on the loop. This is not always the case since the tail may make almost full circle around the loop, but do not pierce the minimal surface spanned on the loop (see Table~\ref{table:highL0}).  Such complex configurations had not been identified by methods that studied intersetions with the minimal surface spanned on the loop~\citep{niemyska2016complex}.



\subsection*{Classification of lasso protein structures and entangled but unpierced loops}

In this section we describe some methods to classify proteins with lassos based on the Gauss linking integral. We propose a precise classification of loop-tail pairs having distinct linking motifs presented by the GLN fingerprints (Fig.~\ref{fig:matrices}).  This is based on three positive real numbers $t_L,t_{L+},t_{LS}$ (for instance $t_L, t_{L+} \approx 0.6, t_{LS} \approx 1.5$), as follows:  \\
$\bullet$ {\bf gL$_0$} - if $max|GLN|\leq t_L$, 
$\bullet$ {\bf gLS} - if $max|GLN|>t_{LS}$; \\
In the all next three cases we demand that $max|GLN|\in (t_L,t_{LS}]$, and: \\
$\bullet$ {\bf gL$_1$} - if exactly one value of $maxGLN$ and $-minGLN$ is greater than $t_L$,  \\
$\bullet$ {\bf gL$_{2+}$} - if both values $maxGLN$ and $-minGLN$ are greater than $t_{L}$ and $|whGLN|\leq t_{L+}$, \\
$\bullet$ {\bf gL$_{3+}$} - if both values $maxGLN$ and $-minGLN$ are greater than $t_{L}$ and $|whGLN|> t_{L+}$.\\
One can consider whole triple consisting of a loop and two tails:  if one of the tails is classified as gL$_0$, then we say that the triple is of the type of the second tail;  if both tails are classified in different way than gL$_0$, we say that the triple is of the type {\bf gLL}.

\begin{figure*}
\centering
  \includegraphics[width=.75\linewidth]{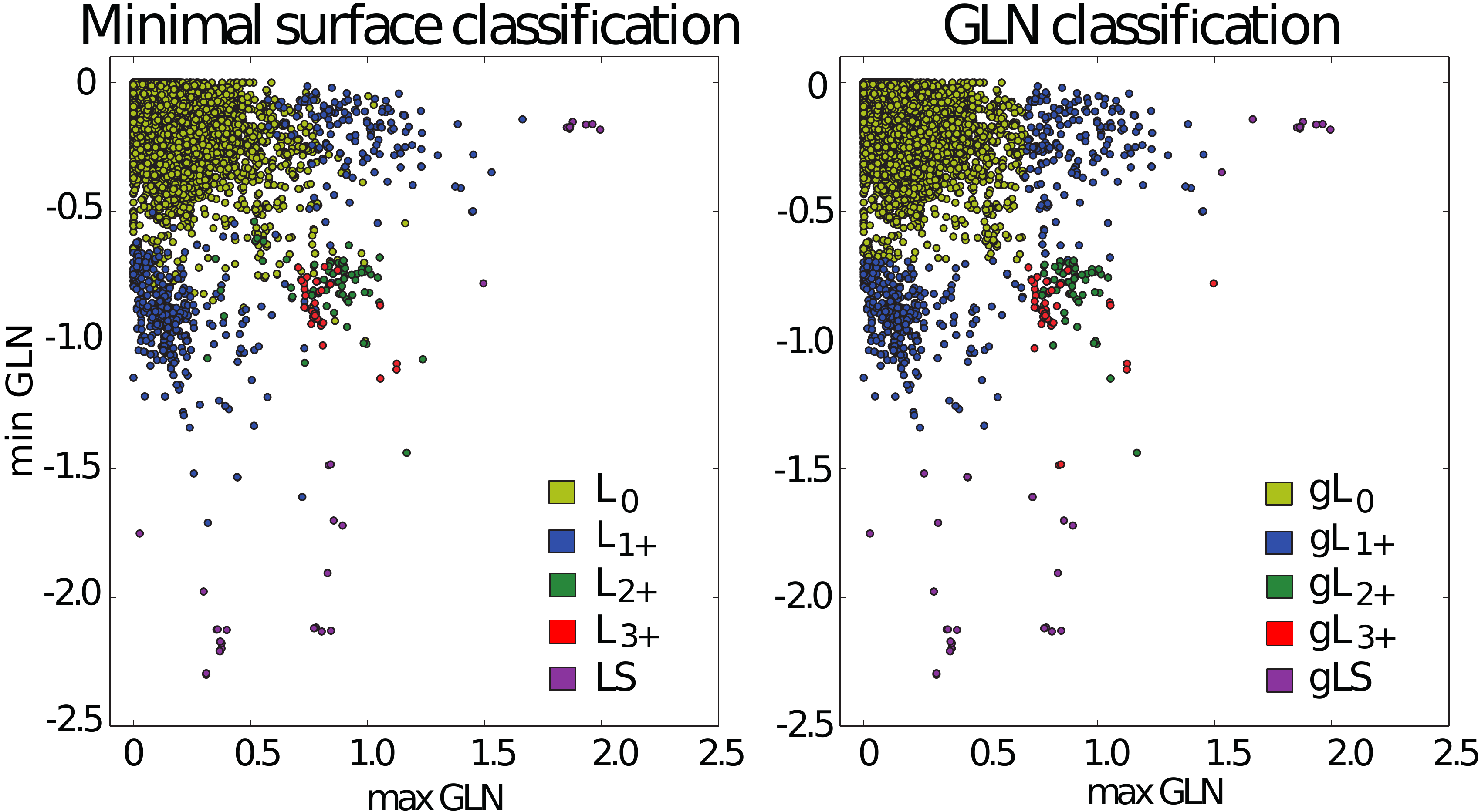}
\caption{Classification of proteins with closed covalent loop based on the minimal surface technique (left) and GLN technique (right). As much as 98\% of structures are classified in an analogous way by both techniques (corresponding points are colored in the same way on both plots). However, on the right, plot types are divided more regularly since the corresponding classification is based only on the GLN values. To differenciate between the types gL$_{2+}$ and gL$_{3+}$ on the plots (green and red dots, respectively) one needs the third coordinate - $whGLN$ value.}
  \label{types_2D}
\end{figure*}

Let L$_{2+}$ denote the sum of types L$_{2n}$ for any natural $n\geq 1$ (in proteins we have found so far examples of L$_2$, L$_4$ and L$_6$, see~\cite{dabrowski2016lassoprot}).  Let L$_{3+}$ denote the sum of types L$_{2n+1}$ for any natural $n\geq1$ (in proteins we only know examples of L$_3$). 
We found that it is possible to choose particular values of $t_{L}, t_{L+}, t_{LS}$ (i.e. $t_{L}=0.69, t_{L+}=0.6, t_{LS}=1.55$) such that as much as $98\%$ of loops are classified in an analogous way by both the techniques of minimal surfaces and the GLN as shown in the Fig.~\ref{types_2D} (see Supplementary Information Fig.~5 for detailed comparison). 
Most of the remaining $2\%$ of loops are structures with intriguing properties that were not recognized before \cite{niemyska2016complex}. We split them into the three groups. 


The first group consists of proteins in which the minimal surface spanned on the loops are not pierced but the tails strongly wind around the loop, or the surfaces spanned on loops are twisted and wind around the tails. When the loop is twisted it appears that there is not enough space to thread the tail through the loop although it is composed of more than 100 amino acids. There are only $15$ such proteins among the set of non-redundant chains of a length lower than $500$ amino acids (see Table~\ref{table:highL0}), with $max|GLN|>0.69$ and no piercings. 
One can ask how does this type of entanglement influence the free energy landscape of the protein 
in oxidizing conditions? We speculate that, in this case, some part of the configurational space is excluded from protein backbone exploration during folding. Unwanted threading will have to backtrack thereby slowing down folding or even leading to missfolding. 

The second group contains proteins with high $|GLN|$ values and the closed loops that are pierced by the tails, but, in minimal surface technique, these piercings are interpreted as being too shallow and are reduced, i.e. they are not taken into account. (Generally, this is a reasonable approach since, for instance, all helices that are crossing surfaces usually do cross them at least three times on a short distance.  We wish to interpret this as simply one meaningful crossing.  However, it is not an easy problem to distinguish shallow crossings from relevant ones (see Supplementary Information  Fig.~6) and the~parallel analysis of GLN matrices may be very helpful in recognizing which reductions are justified or are spatially reasonable.)

The third group consists of structures with low $max|GLN|$ value but with tails piercing the minimal surface spanned on the loops. There are only 9 such loops (0.01\% of the analyzed data set), see Supplementary Information, Table 1. 
(These structures have $max|GLN|\leq0.6$ and no examples with $max|GLN|<0.5$ (see Table~1 in Supplementary Information). With a detailed analysis, we found that in some structures the GLN value is low because the piercing segment lies in the plane of the loop - i.e. is quite "shallow".)
	

\begin{table}[!th]
\caption{``Entangled`` proteins without piercing through a covalent loop closed by a disulfide bridge. 
Based on loops from non-redundant chains of a length lower than $500$ amino acids, which are not pierced, but have $max|GLN|>0.69$.}
\label{table:highL0}
\begin{center}
\begin{tabular}{c | c | c | c }
\textbf{Protein} & \textbf{Loop} & \multirow{2}{*}{\textbf{Tail}} & \multirow{2}{*}{\textbf{Max$|$GLN$|$}} \\ 
\textbf{(chain)} & \textbf{range} & &   \\ \hline \hline
2bb6  (A) & 98-294 & N  & 0.99      \\  
1ece (A) & 34-120 & N  & 0.91       \\ 
4e9i  (C) & 53-135 & C   & 0.87      \\ 
4df0 (A) & 148-198 & N & 0.83       \\  
3vv5  (A) & 97-235 & N  & (-) 0.79  \\ 
2pmv (A) & 85-270 & N   & 0.78     \\ 
4m82 (A) & 275-397 & N & 0.78      \\ 
1uhg (A)& 73-120& C  & 0.77      \\ 
4wtp (A) & 218-264 & N & 0.76 \\
2b34  (A) & 20-114 & N  & 0.74  \\
2x5x  (A) & 36-85 & C    & 0.74  \\
5acf  (A) & 41-167 & C   & 0.73 \\
1qfx (A) & 52-368 & C  & (-) 0.72 \\
5fzp (A) & 12-72 & C    & (-) 0.72\\
2dw2 (A) & 308-388 & N & 0.70 \\
\end{tabular}
\vspace{0.1cm}\\
\end{center}
\end{table}


\subsection*{Unique biophysical features of lasso proteins}

\begin{figure*}
\centering
  \includegraphics[width=.99\linewidth]{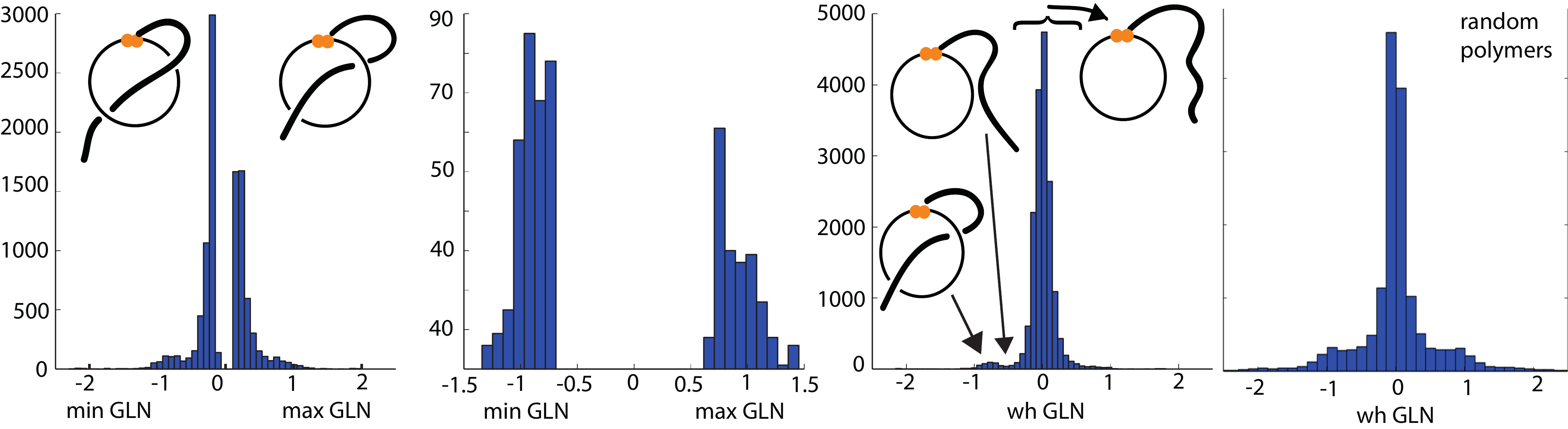}
  \caption{Distribution of $maxGLN$, $minGLN$ and $whGLN$ values based on the 13,320 loops closed by disulfide bridges. Panels A,B,C indicate that there are more negative GLN values than positive ones in proteins. A) Histogram of all $maxGLN$ and $minGLN$ values that are greater than $0.15$ or lower than $-0.15$, and 53\% of them are negative. B) Histogram of all $maxGLN$ or $minGLN$ values (only greater value - in the sense of absolute value - from each pair is taken into account here) from the loops of gL$_1$ type - over 63\% of them are negative. 
C) Histogram of all $whGLN$ values in the analyzed dataset revealing the local minimum around the value $-0.5$. D)  Histogram of $whGLN$ for random polymers.
}
  \label{fig:hist_max}
\end{figure*}

An analysis of the statistics concerning GLN reveals interesting features from the biological point of view. First of all, the windings in the negative direction occur significantly more often than those in the positive direction. For example, among the loops of gL$_1$ type over 63\% have a negative GLN value (see Fig.~\ref{fig:hist_max}, panel B). However, a detailed analysis of basic physico-chemical properties (a~type of amino acids, type of disulfide bridge \citep{bulaj2005formation}) does yet not provide an explanation of this difference. 

The histogram of all $whGLN$ values reveals a noticeable depression around the value $-0.5$ (see Fig.~\ref{fig:hist_max}, panel C). This shows that there are only a few tails that come close to the loop but are not pierced through it. In the case of the random polymers with the same size of the loop and tails, such behaviour is not observed (see Fig.~\ref{fig:hist_max}, panel D). This implies that the depression in proteins distribution arises from a specific side chain interaction which makes contacts outside the loop or, if they are close enough, to the loop whose the minimal surface spanned on the loop they would pierce. 
%

\begin{figure*}
\centering
 \includegraphics[width=1.1\linewidth]{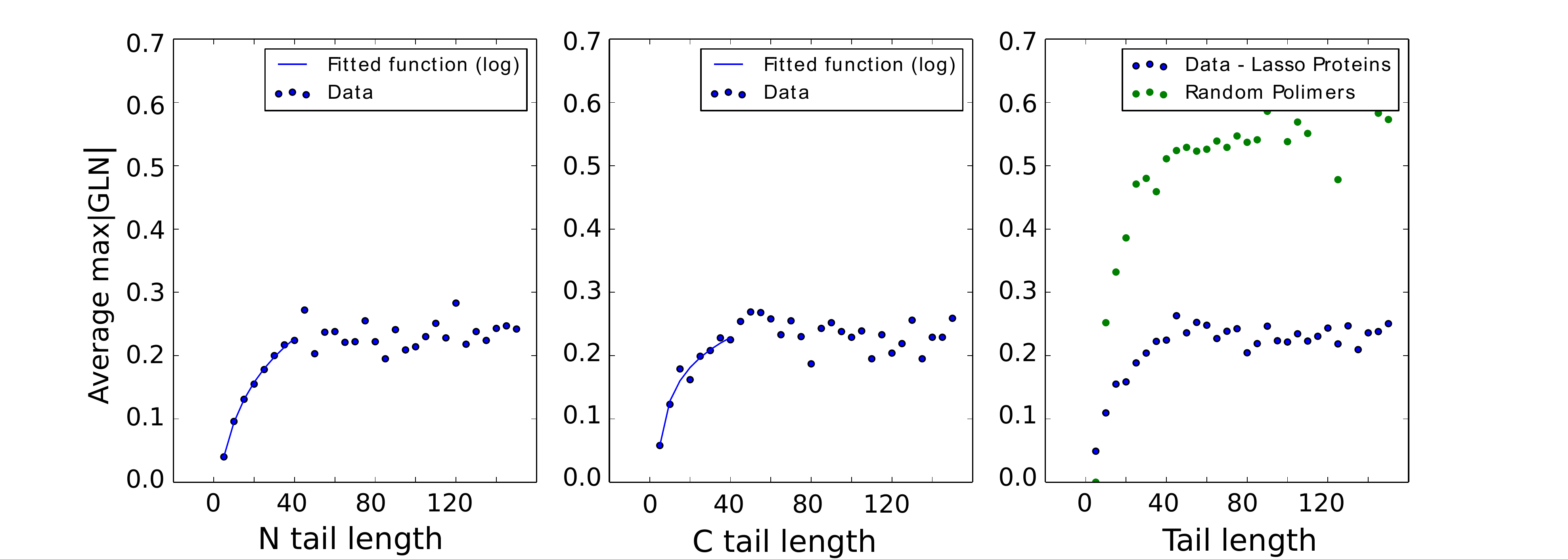}
\caption{Left and middle panels: average $max|GLN|$ values for different lengths of N and C-tails, respectively - first they grow logarithmically, then they become more or less constant, equal to about $0.25$. Right panel: comparison of average $max|GLN|$ values for different lengths of tails in proteins (N and C-tails counted together) and for random polymers. The plot reveals a similar pattern but with much higher GLN values in polymers, stabilizing around $0.55$.}
\label{fig:tailsloops}
\end{figure*}

Considering the lengths of loops and tails we find that the average value of $maxGLN$ depends logarithmically on the length of a tail, up to a length of around 40 amino acids. 
Next, $maxGLN$ saturates and remains stable around the value 0.25 (0.55 for polymers) (see Fig.~\ref{fig:tailsloops}). 




Finally, the analysis of B-factors (the temperature factor) shows that in~chains with short loops amino acids
for which $|GLN|$ between the loop and the tail's fragment from begining to the amino acid is the highest, 
have higher B-factors than average ones. Moreover, amino acids for which $|GLN|$ between the loop and the unit segment corresponding to the amino acid is the highest (often those segments pierce the minimal surface spanned on the loop) -- have significantly lower B-factors, lower even than amino acids creating cysteine bridges. For all loops the tendency is similar, however a little bit less strong (see Table~\ref{table:Bfactor}). This suggests that the parts of tails piercing the loops spanning surfaces are more stable, while the parts of tails between bridges and crossings fluctuate more. 
This is in agreement with available experimental data for lasso type polypetides \citep{zimmermann2013astexin}. 

\begin{table}[!th]
\caption{Correlation between GLN values (of unit segments of tails and whole loop) and B-factors for corresponding amino acids in lasso proteins. Second column: proteins with loops consisting of less than 50 amino acids are taken into account. Third column: all loops.}
\label{table:Bfactor}
\begin{center}
\begin{tabular}{ r | c | c  }
Average $B$-factor for amino acids -- & Short loops & All loops   \\ \hline \hline
all $<{\it amino~acids}>$ & 31.0 & 28.3    \\  \hline
creating bridges & 29.7 & 28.4    \\  \hline
for which GLN between the loop & \multirow{3}{*}{34.2} &  \multirow{3}{*}{30.7}\\ 
and the tail's fragment from&  & \\
beginning to them is the highest  &  & \\  \hline
for which GLN between the loop  & \multirow{2}{*}{26.4} &  \multirow{2}{*}{23.7}\\
and the unit segment is the highest &  &   \\ \hline
\end{tabular}
\vspace{0.1cm}\\
\end{center}
\end{table} 

\begin{figure}
\centering
  \includegraphics[width=.9\linewidth]{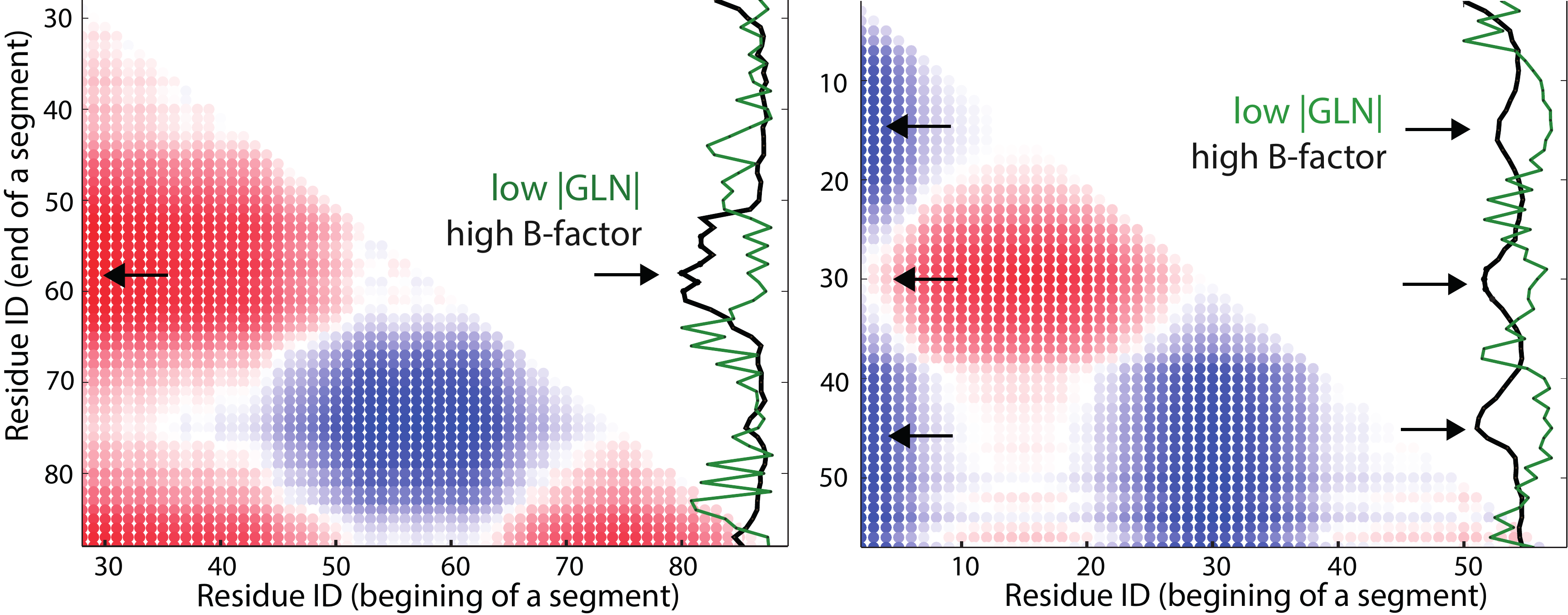}
\caption{Correlation between GLN values and B-factors shown in the GLN matrices for proteins (left: pdb id 4ors, with the loop closed by amino acids 89-186, right: pdb id 2ehg, with the loop 58-145; matrices are for N-terminals), both of gL$_3$ type. On the right edge of the matrix, B-factors are in black and $|GLN|$ values between unit segments and whole loop are in green.  Note that when a local $|GLN|$ is high it usually means that the tail is just winding around the loop, which results in color changes on the left edge of the matrix. When local $|GLN|$ is low, the tail is often far from the loop, not winding around it as significantly at that location. }
\label{fig:bfactor}
\end{figure}

The strong correlation between GLN values of unit segments and whole loop, and B-factors for corresponding amino acids is clearly visible in Fig.~\ref{fig:bfactor}. High B-factors correlate with low $|GLN|$ values and inversely - high $|GLN|$ values correlate with low B-factors. This again suggests that pieces of the tail winding around the loop are more stable that the other segments of the tail. 

\subsection*{Applications of the GLN fingerprint}

Understanding the mechanism by which proteins fold to their native structure is a central problem in protein science \citep{best2013native}. In the case of a majority of proteins, native contacts are sufficient to drive the folding of the protein
\citep{bryngelson1995funnels,wolynes1995navigating,thirumalai2010theoretical} since their free energy landscape is  minimally frustrated \citep{wolynes2005recent}. The fraction of native contacts, called Q, was shown to be a good reaction coordinate to study the folding mechanism for a majority of proteins \citep{best2013native}. However, in the case of proteins with non-trivial topology (e.g. the smallest knotted protein MJ0366 \citep{bolinger}), Q merely represents the progress of folding \citep{dabrowski2015prediction}.

\begin{figure*}
\centering
  \includegraphics[width=1.\linewidth]{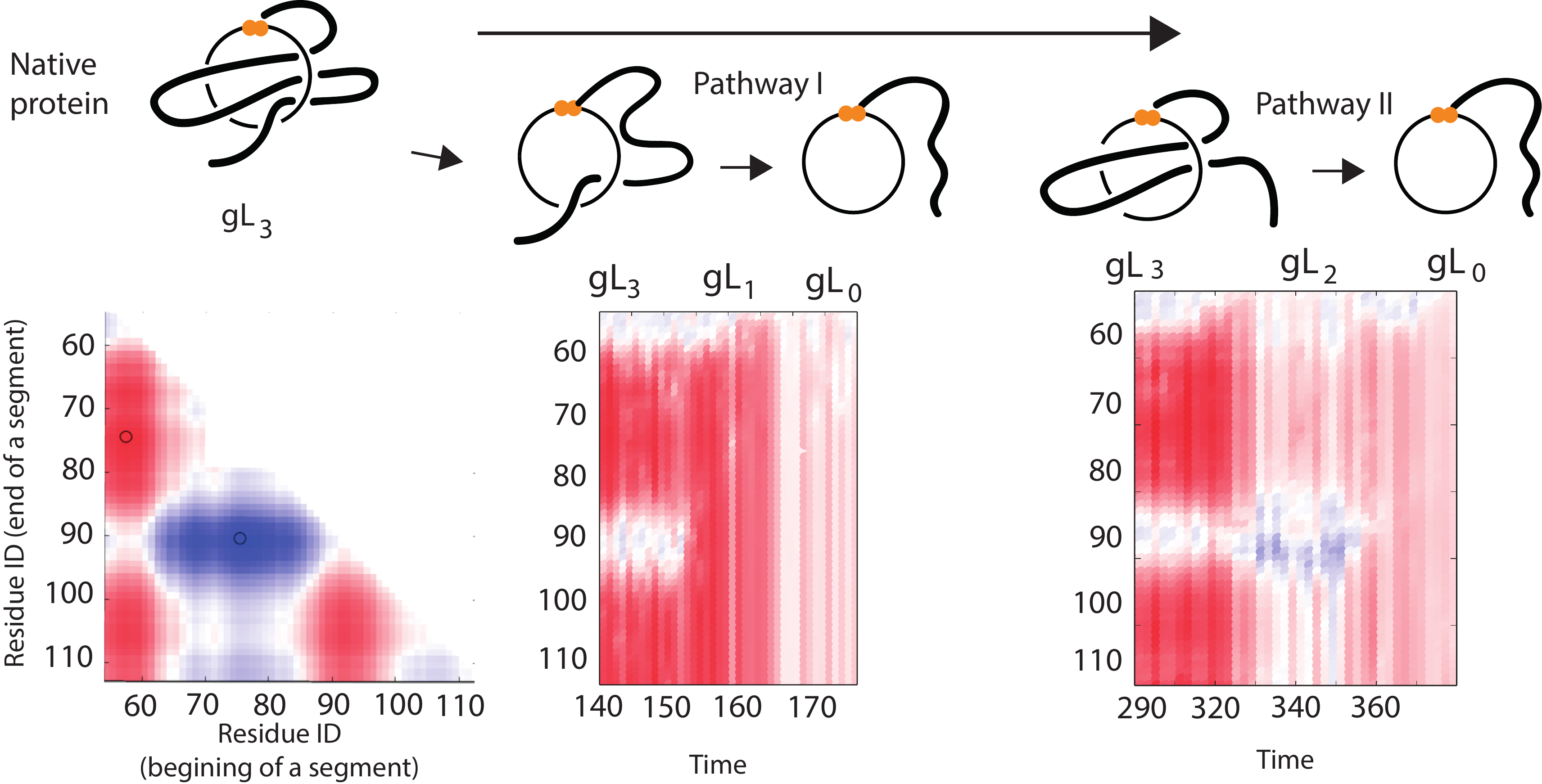}
  \caption{Example of two topologically different unfolding routes identified with GLN method for the ribonuclease U2 (pdb ID 3agn)
with $gL_{3}$ motif (the closed loop is pierced three times). Left panel: the GLN matrix at the native conformation.
Middle panel: visualization of unfolding via unthreading internal loop toward $gL_{1}$ motif, next single unthreading to trivial topology. Each column of this matrix corresponds to the single time frame in the simulation and represents left edge of the GLN matrix for this frame. Right panel: untying to $gL_{2}$ geometry, next untying via slipknot motif to $gL_{0}$.} 
  \label{folding}
\end{figure*}

Next, we show that the GLN values and the GLN fingerprint can reveal information, hidden from Q, about the topology 
based on unfolding pathways simulated with a structure based model \citep{smog}. In fact, in the case of the ribonuclease U2 protein with the $gL_{3}$ motif (the loop is pierced three times), GLN values reveal an ensemble of the transition states composed of at least two unfolding pathways: via the slipknot topology \citep{sulkowska2009dodging,noel2010slipknotting} or direct unthreading (see Fig.~\ref{folding}).  Moreover, superposition of the fingerprints over the time shows how the protein backbone travels through the available conformation space. 
The same technique can be applied to reveal untying of even more complex topologies such as the supercoling motif $gLS$
(one tail winding around the loop and piercing it two or more times from the same site).  The unfolding pathway for a protein with $gLS_3$ is shown in Supplementary Information Fig.~7.

The application of the GLN is not limited to studying lasso proteins or proteins with links \citep{dabrowski2017topological}. Since the GLN measures mutual entanglement its fingerprint is different for 
``the same`` protein with two topologies -- unknotted and knotted (see Supplementary Information Fig.~8) \cite{notatka3}.
Furthermore, the pattern of the GLN fingerprint can be used to identify the  type of secondary structures of the protein which are usually visible via a contact map. 
Note, that the shape of the contact map depends on the cutoff distance used to determine physical contacts while GLN does not depend on additional parameters.  Moreover, sign of GLN (blue or red color on the matrix) indicates the "direction of contact", i.e. from this it can be deduced on which side the fragments of protein chain being in contact pass each other (for more details see Supplementary Information Fig.~8, Fig.~9). Thus, the GLN fingerprint of a native conformation can be used as a reference value for a reaction coordinate in studying the folding pathways of protein. 

\section*{Discussion and conclusions}

We have shown that the GLN method is a significantly faster technique to detect entanglement in proteins with closed loops in the comparison with the methods which rely on minimal surfaces spanning the covalent loops ~\citep{niemyska2016complex}.
The method also reveals much more information about the geometry of chains with lassos which may lead to the new biological and chemical discoveries. 
However, the algorithm based on the surfaces has the advantage of giving precise information about the exact residues that cross the spanning surface which may lead to an important insight from the biological point of view.  We believe both approaches can compliment each other and, together, help focus study on important features of the protein.

The GLN fingerprint can also be used to compare proteins e.g. during CASP or CAPRI competition. Indeed, it can be pushed further, so that the GLN fingerprint provides a powerful tool to be used to improve already very successful deep learning algorithms used to  predict tertiary and quaternary structure of proteins via image recognition \citep{gao2019destini}.

The present method can be applied to any structure in which a loop and tail can be defined. Apart from the cysteine bridge loops investigated here, a loop can be formed, among others, by a salt bridge, by a hydrogen bond, or by ions.  An example of the last case is the human transport protein (PDB code 1n84), with the loop closed by Tyr95-Fe339-Asp63 interaction whose spanning surface is pierced by C-terminal tail (Thr250)~\citep{Dabrowski2018thesis} thus forming lasso of gL$_1$ type.

Moreover, one can apply GLN approach to study entanglement between two structures none of which are closed loops. Lately new algorithm, GISA, was proposed to study local entanglement in protein chains and other biopolymers  \citep{gronbaek2019gisa}. The algorithm computes Gauss integrals between many pairs of quite short fragments of chain and finds rare invariant values. It can be helpful in search for knots, links and highly entangled configurations not previously described as well.
Furthermore since this approach is much faster than other linking invariants it will provide a very useful technique to study loops in a single chromosome as well as chromosome entanglement in the cell \citep{sulkowska2018knotgenome,niewieczerzal2019defining}. Current methods allow one to describe single chromosomes with high resolution (thousands of beads). This number is already an order of magnitude bigger than the typical length of the protein.


\section*{Materials and Methods}

{\bf Gaussian linking number.} A definition of linking number between two closed curves $\gamma_1$ and  $\gamma_2$ in $3$ dimensions is given by the Gauss double integral,

\begin{equation}
GLN\equiv \frac{1}{4\pi}\oint_{\gamma_1}\oint_{\gamma_2}\frac{\vec{r}^{(1)}-\vec{r}^{(2)}}{|\vec{r}^{(1)}-\vec{r}^{(2)}|^3}\cdot (d\vec{r}^{(1)}\times d\vec{r}^{(2)}),
\end{equation}
where $\vec{r}^{(1)}$ and $\vec{r}^{(2)}$ are positions of two curves. Gauss proved that, for closed oriented curves, this integral is always integer, is an invariant up to isotopy, and measures how many times one curve winds around the second one. In the protein case chains become collections of points, i.e., positions of C$\alpha$ atoms $\{ \vec{r}_1^{(k)}, \vec{r}_2^{(k)},\ldots  \vec{r}_{N_k}^{(k)} \}$, for the chains of the length $N_k$, $k=1,2$. The integrals may be replaced by sums over segments $d\vec{R}_i^{(k)}=\vec{r}_{i+1}^{(k)}-\vec{r_i}^{(k)}$, for which we use the midpoint approximation $\vec{R}_i^{(k)}=(\vec{r}_{i+1}^{(k)}+\vec{r_i}^{(k)})/2$. We can replace the requirement of having oriented closed loops by oriented open arcs giving a real value as a measure of linking rather than an integer. We can then perform the double Gauss discrete integral over the open chains,
\begin{equation}
GLN\equiv \frac{1}{4\pi}\sum_{i=1}^{N_1-1}\sum_{j=1}^{N_2-1}\frac{\vec{R}_{i}^{(1)}-\vec{R}_j^{(2)}}{|\vec{R}_i^{(1)}-\vec{R}_j^{(2)}|^3}\cdot (d\vec{R}_i^{(1)}\times d\vec{R}_j^{(2)}).
\end{equation}

\textbf{Note, one can simply employ the Banchoff method on the open chain to explicitly calculate this integral \citep{banchoff}.}\\
Let us denote 
\begin{equation}
G(i,j):=\frac{\vec{R}_{i}^{(1)}-\vec{R}_j^{(2)}}{|\vec{R}_i^{(1)}-\vec{R}_j^{(2)}|^3}\cdot (d\vec{R}_i^{(1)}\times d\vec{R}_j^{(2)}),
\end{equation}
$i\in\{1\ldots N_1-1\},j\in\{1\ldots N_2-1\}$, and consider a pair of a tail of a length $N_1$ and a loop of a length $N_2$.
We calculate and then analyze four main values for each pair of a loop and a tail:
\begin{itemize}
\item $whGLN$: value of the Gauss double integral between a loop and whole tail,
\begin{equation}
whGLN = \frac{1}{4\pi}\sum_{i=1}^{N_1-1}\sum_{j=1}^{N_2-1}G(i,j);
\end{equation}
\item $minGLN$ ($maxGLN$): minimum (maximum) value of the Gauss double integral between a loop and any fragment of a tail,
\begin{equation}
minGLN = \min_{\substack{
   k,l\in\{1...N_1-1\}, \\
  k< l
  }} \frac{1}{4\pi}\sum_{i=k}^{l}\sum_{j=1}^{N_2-1}G(i,j);
\end{equation}
\item $max|GLN|=\max\{maxGLN,-minGLN\}$.
\end{itemize}
Additionaly for each triple of a loop and two tails we considered $max2|GLN|$, which is a maximum of $max|GLN|$ for both tails.

{\bf Protein dataset.} We use the set of 5,106 non-redundant proteins with at least one bridge from LassoProt database~\citep{dabrowski2016lassoprot}, March 2016. By {\it non-redundant} we mean sequence similarity is lower than 35\%, including X-ray, NMR, CEM structures and proteins with unresolved parts. We chose only one chain from each protein and identified 13,320 covalent loops in a total. 
This dataset includes 1,276 chains with unresolved parts which were reconstructed with Gaprepair \citep{jarmolinska2018gaprepairer}
based on Modeller~\citep{webb2014protein}. For details see Supplementary Information file.

{\bf The minimal surface method and molecular visualization.}
The surface is approximated by a discrete triangulation as described in  \citep{niemyska2016complex,dabrowski2016lassoprot}. 
To distinguish structures with the same number of piercings but where the way  he minimal surface spanned on the loop is pierced is different, 
an orientation of the surface spanned on the disulfide loop was introduced. 
Two piercings may occur if the tail pierces the loop in one direction and then the inverse (the $L_2$ structure), or pierces it twice in the same direction, winding around the loop (the $LS_2$ structure). 
Additionally Pylasso \citep{gierut2017pylasso} and PyLink \citep{Pylink} plugin for PyMOL were used to facilitate analysis and perform Molecular graphics. 

{\bf Molecular dynamics simulation.}
The kinetics data were obtained based on a coarse-grained model and conducted using the Gromacs package with SMOG software \citep{smog} employing parameters from \citep{sulkowska2008selection}.

{\bf Random lassos sampling.}
Phantom lassos (polymers deprived of any interactions and volume) were created by connecting phantom loops and phantom tails. Phantom loops were created as equilateral polygons using the dedicated algorithm \citep{cantarella2016fast} and tested earlier in the \citep{dabrowski2019statistical}.



\section*{Acknowledgments}
\begin{small} 
The authors would like to thank Szymon Niewieczerzal, Bartosz Gren for help with running simulations,
Eleni Panagiotou, Pawel Dabrowski-Tumanski for useful discussions. This work was financed from the budget of Polish Ministry for Science and Higher Education Grant [\#0003/ID3/2016/64 Ideas Plus] to JIS, and University of Warsaw [\#501-D313-86-0117000-03] to WN.
\end{small}

\section*{Author Contribution}
\begin{small}
J.I.S., K.C.M. and W.N designed the work, W.N. and J.I.S performed the work and wrote the paper.
\end{small}
\section*{Additional information}

\begin{small}
Supplementary Information is attached.
\textbf{Competing financial interests:} The authors declare no competing financial interests.
\end{small}

\bibliography{biblio52}

\begin{thebibliography}{10}
\bibitem{notatka3}
The protein from si (pdb id 1oy5) was crystallized as unknotted, however, it is
  expected that it is knotted since it was shown that the knotted topology is
  strictly conserved for homologous proteins \citep{sulkowska2012conservation},
  the knotted homological protein (pdb id 4yqd) was used to model the knotted
  version of this protein.
\end{thebibliography}

\end{document}